\documentclass
[10pt,superscriptaddress,twocolumn,showpacs,letterpaper]{revtex4}%
\usepackage{amsfonts}
\usepackage{amsmath}
\usepackage{amssymb}
\usepackage{graphicx}
\usepackage{graphics}
\usepackage[usenames]{color}%
\providecommand{\U}[1]{\protect\rule{.1in}{.1in}}
\input epsf

\begin{document}
\title{Three body recombination of ultracold dipoles to weakly bound dimers.}
\author{Seth T. Rittenhouse}
\affiliation{ITAMP, Harvard-Smithsonian Center for Astrophysics, Cambridge, Massachusetts
02138, USA}
\author{Christopher Ticknor}
\affiliation{ITAMP, Harvard-Smithsonian Center for Astrophysics, Cambridge, Massachusetts
02138, USA}
\affiliation{ARC Centre of Excellence for Quantum-Atom Optics and Centre for Atom Optics
and Ultrafast Spectroscopy, Swinburne University of Technology, Hawthorn,
Victoria 3122, Australia}
\affiliation{Theoretical Division, Los Alamos National Laboratory, Los Alamos, New Mexico
87545, USA}
\date{\today}

\begin{abstract}
We use universality in two-body dipolar physics to study three-body
recombination. We present results for the universal structure of weakly bound
two-dipole states that depend only on the s-wave scattering length ($a$). We
study threshold three-body recombination rates into weakly-bound dimer states
as a function of the scattering length. A Fermi Golden rule analysis is used
to estimate rates for different events mediated by the dipole-dipole
interaction and a phenomenological contact interaction. The three-body
recombination rate in the limit where $a\gg D$ contains terms which scale as
$a^{4}$, $a^{2}D^{2}$ and $D^{4}$, where $D$ is the dipolar length. When $a \ll D$, the three-boby recombination rate scales as $D^4$.

\end{abstract}

\maketitle

Universality has become increasingly important and apllicable in ultracold
physics. Two remarkable examples are: strongly interacting fermions
\cite{ho,drummond,fermions} and few-body systems with large scattering
lengths. Three-body recombination, $A+A+A\rightarrow A_{2}+A+KE$, is of
special interest in an ultracold gas. It usually is the limiting process
determining the lifetime of the gas in the absence of twp body losses.
Three-body recombination has also proven to be an important probe in
understanding the basic quantum mechanical nature of universal few-body
systems \cite{review}. The experimental observation of universal 3- and 4-body
loss features \cite{grimm,lens,rice}, associated with a class of three
body-bound state predicted by Efimov \cite{efimov}, is one of the major
success stories in this field. These few body complexes have found great
utility because of the control over the 2 body scattering length with a
magnetic Fano-Feshbach resonance \cite{feshbachRMP}. In a magnetic field, the
last bound state can be tuned through the scattering threshold giving complete
control over the scattering length. When $a$ is much greater than the range of
the short range interaction, many properties become independent of the details
of the two-body interaction. For example, the three-body recombination takes
on universal scaling behavior of $a^{4}$ in this regime \cite{fed}.

The powerful idea of universality can be exploited to study complex systems.
Now ultracold polar molecules have emerged \cite{gspm,chem}, and these systems
will have a new type of collisional control \cite{collisions}. This control
emerges in the form of the dipole-dipole interaction, $V_{dd}=d^{2}%
{[1-3(\hat{z}\cdot\hat{r})^{2}]/r^{3}}$ where $r$ is the inter-particle
separation vector and $d$ is the induced dipole moment along an external field
axis ($\hat{z}$). The collisional control can be achieved through the induced
dipole moment $d$ which is determined by the strength of the electric field.
As the electric field is increased, there is a series of s-wave dominated
resonances \cite{You,CTPR,roudnev}. These resonances occur when the system
gains a weakly-bound long-range state. The short-range bound states structure
of the interaction is largely independent of the external fields strength.
These threshold resonances could be used to control scattering length, but the
importance of these resonances will go further. The interaction is a long
range and anisotropic which will mediate fundamentally novel quantum behavior.
What is striking about this interaction is that the length scale describing it
can be incredibly large, many orders of magnitude larger than the short range
interaction. The dipolar length scale is $D=\mu_{2b}d^{2}/\hbar^{2}$ and
$\mu_{2b}$ is the reduced mass of a two body system. This large length scale
can be used to create a highly correlated system.

The two-dipole scattering cross section has been demonstrated to be universal
\cite{universal,roudnev,NJP}. In this work we extend the concept of
universality in two-body dipolar physics and apply it to three-body
recombination. We present results showing the universal structure of weakly
bound two-dipole states that depend only on $a$. Using this, we then study
threshold three-body recombination rates into weakly-bound dimer states as a
function of $a$.

To start, we study the universality of the dipolar threshold resonances. These
universal dipolar collisions depend only on the s-wave scattering length. To
tune the scattering length, we vary the short range position of a hard wall
boundary condition \cite{roudnev}. As the inner wall's postition is decreased
the system has a series of threshold resonances; in a real system this mimics
increasing the electric field and the molecule's polarization. In this work we
only study threshold resonances, the wide s-wave dominated resonances. There
are other resonances which are not s-wave dominated, but these will be narrow
and rare \cite{roudnev}.

\begin{figure}[ptb]
\vspace{3mm}
\centerline{\epsfysize=80mm\epsffile{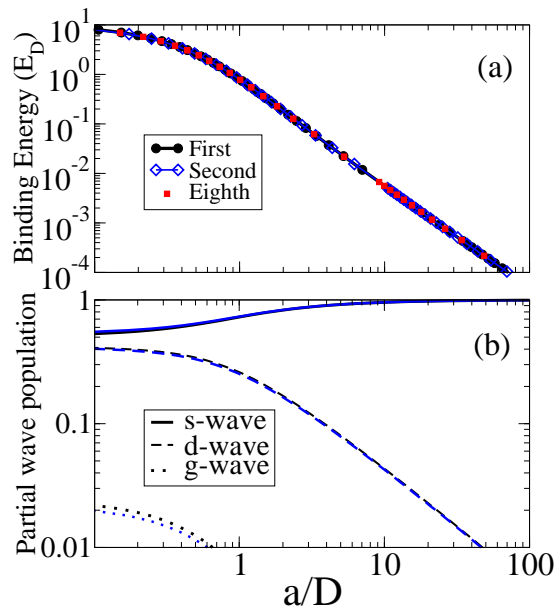}}\caption{(Color Online)
{Universal properties of 2 body dipolar scattering}. (a) The binding energy
(symbols with curves) of the dipolar dimer as a function of the s-wave
scattering length. for the first (black circles), second (blue diamonds), and
eighth (red square) resonance. (b) The partial wave molecular populations are
shown for the s-wave (solid), d-wave (dashed) and g-wave (dotted) populations
for the first (black) and second (blue) resonance. }%
\label{binding}%
\end{figure}

Examples of universal two-body scattering are well known, for instance when
the scattering length is large and positive, large dimers form with binding
energies $E_{b}=\hbar^{2}/(2\mu_{2b}a^{2})$. Weakly bound two-body dipolar
states near an s-wave resonance ($a\gg D$) also exhibit this universal
property. The binding energy for these states is shown in Fig. \ref{binding}
(a) in units of the dipolar energy $E_{D}=\hbar^{2}/\mu_{2b}D^{2}$ as a
function of $a/D$. We have plotted the binding energies for three different
resonances, the first (black circles), second (blue diamonds) and eighth (red
squares) resonance. As expected the three resonances agree very well in the
large scattering length limit, but surprisingly the agreement is very good
even when $a<D$.

To further illustrate universal features, we look at the partial wave
populations as a function of $a$. We have numerically obtained the molecular
wave function: $\psi_{d}(\vec{r}_{12})=\sum_{l}{Y_{l0}(\hat{r}_{12})}f_{d}%
^{l}(r_{12})$ where $l$ is the partial wave. We require that $\sum_{l}N_{l}=1$
where $N_{l}=\langle f_{d}^{l}|f_{d}^{l}\rangle$ is the partial wave
population. Figure \ref{binding} (b) shows the s- (solid), d- (dashed), and
g-wave (dotted) populations for the first (black) and second (blue) resonances
as a function of $a/D$. This figure shows that the partial wave populations
are universal. Additionally it shows that these molecules are mostly s- and
d-wave. For large $a$, the s-wave contribution dominates and is near unity for
large $a$ and $\psi_{d}$ takes on the universal form $r\psi_{d}\propto
\exp\left(  -r/a\right)  $. As $a$ is decreased the d-wave contribution
becomes significant and reaches  $\sim40\%$. Even for very small $a$, the
g-wave, i.e. $l=4$, contribution is at most 2$\%$. The dipolar interaction
conserves parity and this selection rule prevents an s-wave channel coupling
to any odd partial wave. This implies that the present theory applies to only
bosonic or distinguishable dipoles.

There is of course a caveat to universality. The dipole length, $D$, must be
much larger than any short range length scale, such as the depolarization,
\textquotedblleft dipole flip" length scale which is typically 100
$a_{0}$ where $a_{0}$ is the Bohr radius \cite{roudnev2}. At this point the short range physics will become important and the
structure of the bound state will be system dependent. However, these results
are widely applicable because the significant size of the dipolar length
scale. It can easily reach $10^{3}a_{0}$, some can even exceed $10^{5}a_{0}$
for heavy, very polar molecules, such as LiCs.

To find the three-body recombination cross section we begin by considering
three distinguishable dipoles and use a simple Fermi's golden rule based
approach which begins with the differential rate given by%
\begin{equation}
dR=\dfrac{2\pi}{\hbar}\left\vert \left\langle \Psi_{2+1}\left\vert
V\right\vert \Psi_{1+1+1}\right\rangle \right\vert ^{2}\delta\left(
E_{i}-E_{f}\right)  .\label{Eq:FermiGoldenRulediffRate}%
\end{equation}
Here the initial state, $\Psi_{1+1+1}$, is a three-body, box-normalized
plane-wave in Jacobi coordinates. The final state, $\Psi_{2+1}$, consists of a
weakly bound dimer and a box-normalized dipole-dimer plane wave. Because the
dipole-dipole interaction has no diagonal s-wave contribution, we consider a
modified two-body potential in Eq. (\ref{Eq:FermiGoldenRulediffRate}) which
incorporates the zero-range Fermi-pseudo potential to account for s-wave
scattering, i.e.%
\begin{equation}
V\left(  \vec{r}\right)  =\dfrac{d^{2}\left[  1-3\left(  \hat{r}\cdot\hat
{z}\right)  ^{2}\right]  }{r^{3}}+\dfrac{2\pi\hbar^{2}a}{\mu_{2b}}%
\delta\left(  \vec{r}\right)  .\label{Eq:Ef_int}%
\end{equation}
where $a$ is the positive s-wave scattering length near a dipole-dipole
threshold scattering resonance and $\delta$ is the standard Dirac $\delta
$-function. Summing over all final-state energies and dividing by the total
incident particle flux yields the total differential recombination cross
section for three distinguishable dipoles \cite{mottandmassy},%
\begin{align}
\dfrac{d\sigma^{dist}}{d\hat{k}_{2}^{\prime}} &  =\dfrac{\mu_{4b}^{2}%
k_{2}^{\prime}}{4\pi^{2}\hbar^{4}k}\left\vert \int d^{3}\rho_{1}d^{3}\rho
_{2}\psi_{d}^{\ast}\left(  \vec{\rho}_{1}\right)  e^{-i\vec{k}_{2}^{\prime
}\cdot\vec{\rho}_{2}}\right.  \label{Eq:diffcrosssec1}\\
&  \times\left.  V\left(  \vec{r}_{13}\right)  e^{i\left(  \vec{k}_{1}%
\cdot\vec{\rho}_{1}+\vec{k}_{2}\cdot\vec{\rho}_{2}\right)  }\right\vert
^{2}.\nonumber
\end{align}
Here $\vec{\rho}_{i}$ is a mass scaled Jacobi vector given by%
\begin{align*}
\vec{\rho}_{1} &  =\sqrt{\dfrac{\mu_{12}}{\mu_{4b}}}\left(  \vec{r}_{1}%
-\vec{r}_{2}\right)  ,\\
\vec{\rho}_{2} &  =\sqrt{\dfrac{\mu_{12,3}}{\mu_{4b}}}\left(  \dfrac{\vec
{r}_{1}+\vec{r}_{2}}{2}-\vec{r}_{3}\right)  ,\\
\mu_{12} &  =\dfrac{m_{1}m_{2}}{m_{1}+m_{2}};\mu_{12,3}=\dfrac{\left(
m_{1}+m_{2}\right)  m_{3}}{m_{1}+m_{2}+m_{3}},\\
\mu_{4B} &  =\sqrt{\dfrac{m_{1}m_{2}m_{3}}{m_{1}+m_{2}+m_{3}}}.
\end{align*}
In Eq. (\ref{Eq:diffcrosssec1}) $\vec{k}_{2}^{\prime}$ is the outgoing
dipole-dimer wave number with magnitude $k_{2}^{\prime}=\sqrt{2\mu_{4B}\left(
E_{inc}+E_{b}\right)  }$, $\vec{k}_{1\left(  2\right)  }$ is the incoming wave
number associated with the first (second) Jacobi vector, $k=\sqrt{k_{1}%
^{2}+k_{2}^{2}}=\sqrt{2\mu_{4B}E_{inc}}$ is the magnitude of the initial
three-body plane wave with total incident energy $E_{inc}$. It should be noted
that only a single two-body interaction is considered in Eq.
(\ref{Eq:diffcrosssec1}). This is the simplest matrix element that connects
the outgoing dipole-dimer state to the incoming plane wave state.

Expanding the plane waves and dimer-wave function in Eq.
(\ref{Eq:diffcrosssec1}) in spherical harmonics and integrating over the
angular degrees of freedom yields the total distinguishable particle cross
section:
\begin{align}
\sigma^{dist} &  =\dfrac{1024\pi^{2}}{3^{1/4}9\hbar k}\sqrt{mE_{b}%
}\label{Eq:crossec1}\\
&  \times\sum_{L^{\prime}l^{\prime},Ll}\left\vert 4a\delta_{L^{\prime
}l^{\prime}}\delta_{L0}\delta_{l0}\left(  -1\right)  ^{l^{\prime}}%
I_{s}^{l^{\prime}}+DI_{d}^{L^{\prime}l^{\prime},Ll}\right\vert ^{2}%
,\nonumber\\
I_{s}^{l^{\prime}} &  =\int f_{d}^{l^{\prime\prime}\ast}\left(  2R\right)
j_{l^{\prime}}\left(  \sqrt{\dfrac{2\mu_{12,3}E_{b}}{\hbar^{2}}}R\right)
R^{2}dR,\nonumber\\
I_{d}^{L^{\prime}l^{\prime},Ll} &  =\int r^{2}drR^{2}dRj_{L}\left(
\sqrt{\dfrac{\mu_{12,3}}{\mu_{4B}}}k_{2}R\right)  j_{l}\left(  \sqrt
{\dfrac{\mu_{12}}{\mu_{4B}}}k_{1}r\right)  \nonumber\\
&  \times f_{d}^{l^{\prime}\ast}\left(  r\right)  j_{L^{\prime}}\left(
\sqrt{\dfrac{2\mu_{12,3}E_{b}}{\hbar^{2}}}R\right)  \left\langle L^{\prime
}l^{\prime}\left\vert \dfrac{V_{d}\left(  \vec{r}_{13}\right)  }{d^{2}%
}\right\vert Ll\right\rangle _{\Omega}.\nonumber
\end{align}
Here $j_{L}$ is a spherical Bessel funtion, $L$ and $l$ are the initial
angular momentum quantum numbers associated with the first and second Jacobi
vectors respectively, $L^{\prime}$ is the angular momentum of the outgoing
dipole-dimer system, and $l^{\prime}$ is the angular momentum of the dimer. We
note that we have moved to standard Jacobi coordinates in these integrals for
ease of calculations, i.e. $R=\sqrt{\mu/\mu_{12,3}}\rho_{2}$ and $r=\sqrt
{\mu/\mu_{12}}\rho_{1}$. The $\delta_{L0}\delta_{l0}$ in Eq.
(\ref{Eq:crossec1}) is due to the strong suppression all incoming angular
momenta greater than zero in s-wave scattering. The expectation value
$\left\langle L^{\prime}l^{\prime}\left\vert V_{d}\left(  \vec{r}_{13}\right)
\right\vert Ll\right\rangle _{\Omega}$ is only over all angular coordinates.
The cross section in Eq. \ref{Eq:crossec1} was found by summing over all final
outgoing momenta, $\hat{k}_{2}^{\prime}$, and averaging over all incoming
directions $\left\{  \hat{k}_{1},\hat{k}_{2}\right\}  $ as would be
appropriate for a gas phase experiment.

The interaction in Eq. (\ref{Eq:diffcrosssec1}) is most easily evaluated with
a biharmonic expansion \cite{theoryangular}, and it can be written as:
${\frac{Y_{20}(\Omega_{12})}{|\vec{r}_{1}-\vec{r}_{2}|^{3}}}={\frac{1}%
{r_{>}^{3}}}\sum_{kmm^{\prime}}\left(  {\frac{r_{<}}{r_{>}}}\right)
^{k}[Y_{km}(\Omega_{<})\otimes Y_{k+2m^{\prime}}(\Omega_{>})]_{2}^{\left(
0\right)  }$ where $r_{>}$ ($r_{<}$) is the greater (lesser) of $r_{1}$ and
$r_{2}$ and $[Y_{km}\otimes Y_{k+2m^{\prime}}]_{20}$ is a rank 2 tensor formed
by the two spherical harmonics. This can then be evaluated by considering
channels made up of two sets of partial wave, one for both $\hat{r}_{1}$ and
$\hat{r}_{2}$.

With this long range interaction one might expect that all partial waves
contribute equally to the recombination process as is the case in two-dipole
elastic scattering.This does not occur for the three-body inelastic scattering
process. As long as the incoming particles are in the threshold regime,
$E_{inc}\ll E_{b}$, all non-s-wave incoming channels are suppressed. This
means that the final sum can be taken with zero initial angular momentum,
$L=l=0$, greatly simplifying the expression.

The final identical boson cross section $\sigma^{ident}$ is found simply by
multiplying by the Bose enhancement factor, $3!$, and restricting the sum in
Eq. (\ref{Eq:crossec1}) to outgoing channels with the appropriate bosonic
symmetry \cite{esry1999tbr}. The incoming channel is restricted to s-wave only
and already obeys the appropriate symmetry. The recombination rate can be
extracted from the inelastic cross section by multiplying by the
characteristic incoming velocity $v=\hbar k/\mu_{4B}$ \cite{esry1999tbr,fed}:%
\begin{align}
K_{3}^{ident}=\dfrac{\hbar k}{\mu_{4B}}\sigma^{ident} &  =\dfrac
{3!\sqrt{mE_{b}}}{m}\dfrac{1024\sqrt{3}\pi^{2}}{3^{1/4}9}\label{Eq:recrate}\\
&  \times\sum_{L^{\prime}l^{\prime}}\left\vert 4a\delta_{L^{\prime}l^{\prime}%
}I_{s}^{l^{\prime}}+DI_{d}^{L^{\prime}l^{\prime},00}\right\vert ^{2}.\nonumber
\end{align}

In general the integrals in Eq. (\ref{Eq:recrate}) must be performed using
numerically tabulated dimer wave functions $f_{d}^{l^{\prime}}\left(
r\right)  $, but in the case where $a\gg D$ Fig. \ref{binding} (b) shows that
the dipolar wave function is dominated by the s-wave component. Furthermore
the wave function has the universal form for a weakly bound s-wave dimer:
$\psi_{d}\left(  \vec{r}\right)  \rightarrow{e^{-r/a}}/{(r\sqrt{2\pi a})}$.
Inserting this into Eq. (\ref{Eq:recrate}) yields the large $a$ scaling of the
recombination rate:%
\begin{align}
K_{3}^{ident} &  \rightarrow3!\dfrac{32\sqrt{3}\pi^{2}\hbar a^{2}}{m}\left[
a^{2}+\beta D^{2}\right]  ,\label{Eq:recrateunit}\\
\beta &  =\dfrac{64}{15}\left(  2\sqrt{3}-\pi\right)  ^{2}=0.44.\nonumber
\end{align}
The leading order $a^{4}$ term in this is exactly that found in Ref.
\cite{esry1999tbr,metha} for identical bosons with short-range s-wave
interactions with the appropriate scattering matrix element set to $\left(
ka\right)  ^{4}$. The second term is due entirely to the presence of the
dipolar interaction and corresponds to an outgoing d-wave differential cross
section. It should be noted that higher partial wave contributions in the
dimer wave function are present at the $a^{2}D^{2}$ order as well, but this
contribution is two orders of magnitude smaller.

\begin{figure}[ptb]
\vspace{3mm} \centerline{\epsfxsize=3in\epsffile{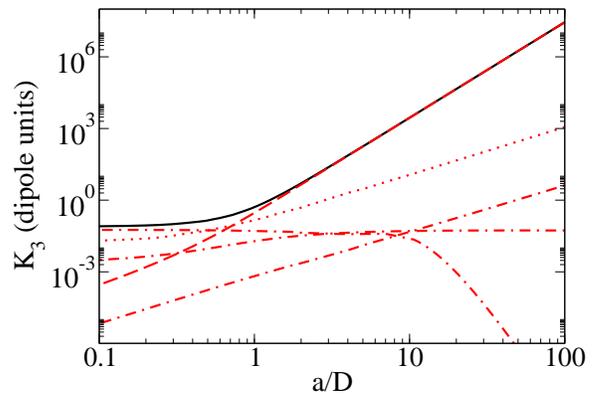}}\caption{(Color
Online) The three-body recombination rate of three identical bosonic dipoles
is shown in dipole units as a function of the scattering length. The total
recombination rate (solid curves) is universal and independent of the
resonance used. The sS (dashed red), dS (dotted red) and higher order
(dot-dashed red curves) partial wave contributions are shown as well.}%
\label{rates}%
\end{figure}

The results of the Fermi gold rule calculations are shown in Fig. \ref{rates}
as a function of scattering length for the first resonance in Fig.
\ref{binding}. The total rates have the expected $a^{4}$ scaling at large $a$
due to the contact interaction, but as $a/D$ is decreased they eventually
flatten out and when $D>a$ the rate is entirely controlled by the dipolar
interaction, scaling as $D^{4}$: $K_{3}^{ident} \rightarrow 0.079 \hbar D^{4}/\mu_{2b}$.
We have compared the recombination rates for several two-body resonances and
have found nearly perfect agreement, to within $\sim1\%$, indicating that the
three-body recombination behavior found here is universal.

The dipolar interaction directly couples the incident channel to final states
with $L^{\prime}=0,2,4,...$ and a molecular state with $l^{\prime}=0,2,...$.
The largest term in the $a>D$ regime, $l^{\prime}L^{\prime}=sS$ (dashed red
curve), comes from the contact interaction and scales as $a^{4}$. The next
largest contribution from the $sD$ channel (dotted red curve) scales as
$D^{2}a^{2}$. Additional contributions from higher order $l^{\prime}L^{\prime
}$ channels are shown (dot-dashed red), but contribute little in the large $a$
regime. However in the $a<D$ regime, the dominant contribution come from the
$l^{\prime}L^{\prime}=dS\text{ and }dG$ terms, the first of which is only
weakly dependent on the scattering length. As one might expect there are two
universal regimes that appear. The first regime appears where the long-range
dipole interactions is dominant and the scattering length is small, i.e. $a\ll
D$, where $K_{3}$ is independent of $a$ and scales as $D^{4}$. The second
regime appears when $a$ is the dominant length scale, $a\gg D$, and the
$a^{4}$ s-wave contribution is dominant over all others.

The Fermi's golden rule based approach is only capable of describing the
scaling behavior of three-body recombination, and does not incorporate more
complex three-body correlations. The recombination rate can have a variety of
interesting resonant behaviors on top of the envelope behavior presented here.
For instance we expect that in the $a \gg D$ regime, a geometrically spaced
set of Efimov minima are likely to appear in the $l^{\prime}L^{\prime}=sS$
outgoing channel \cite{review}.

We have shown the universal structure of weakly bound two-dipole states that
depend only on the scattering length of the system, shown in Fig.
\ref{binding}. This has allowed us to obtain threshold three-body
recombination rates into weakly-bound dimer states as a function of $a$. We
used a Fermi Golden rule analysis to estimate the contributions to the rate
from different events mediated by the dipole-dipole interaction and a contact
interaction. These rates are shown in Fig. \ref{rates}. The individual terms
have many different scaling behaviors. When $a>D$, we find the dominant term
scales as $a^{4}$. When $D>a$ the recombination rate scales as $D^{4}$. The
case of negative scattering length is not addressed in this work, but it is
not unreasonable to expect that the envelope behavior will be similar to that
presented here. More detailed three-body calculations which include extra
resonance structures are the subject of future work. The $|a|\gg D$ regime
might also hold a variety of interesting phenomena such as universal
three-dipole Efimov states.

\begin{acknowledgments}
The authors would like to thank H.R. Sadeghpour for numerous helpful
discussions. Both authors gratefully acknowledge support from the NSF through
ITAMP at Harvard University and Smithsonian Astrophysical Observatory. C.T.
gratefully acknowledges partial support from the Australian Research Council
and Los Alamos Nation Laboratory is operated by Los Alamos National Security,
LLC for the NNSA of the USDoE under Contract No. DE-AC52-06NA25396.
\end{acknowledgments}

\end{document}